\documentclass[preprints,article,accept,moreauthors,pdftex]{Definitions/mdpi} 

\firstpage{1} 
\pubvolume{1}
\issuenum{1}
\articlenumber{0}
\pubyear{2021}
\copyrightyear{2020}
\datereceived{} 
\dateaccepted{} 
\datepublished{} 
\hreflink{https://doi.org/} 

\Title{Single-particle and Collective Structures in Neutron-rich Sr Isotopes}

\TitleCitation{Single-particle and Collective Structures in Neutron-rich Sr isotopes}

\Author{Kamila SIEJA $^{1}$\orcidA{0000-0001-7736-656X}}

\AuthorNames{Kamila SIEJA}

\AuthorCitation{Sieja, K.}

\address{%
$^{1}$ \quad Universit\'e de Strasbourg, IPHC, 23 rue du Loess 67037 Strasbourg, France\\
CNRS, UMR7178, 67037 Strasbourg, France; kamila.sieja@iphc.cnrs.fr}

\corres{Correspondence: kamila.sieja@iphc.cnrs.fr}

\abstract{Neutron-rich Sr nuclei around $N=60$ exhibit a sudden shape transition from spherical
ground state to strongly prolate-deformed. Recently, a lot of new insight into the structure
of Sr isotopes in this region was gained through experimental studies of excited levels,
transitions strengths and spectroscopic factors. In this work, a ``classic'' shell-model
description of strontium isotopes from $N=50$ to $N=58$ is provided, using a natural valence space outside
the $^{78}$Ni core. Both even-even and even-odd isotopes are adressed. In particular, spectroscopic
factors are computed to shed more light on the structure of low-energy excitations and their evolution
along the Sr chain. The origin of deformation at $N=60$ is commented in the context of the
present and previous shell-model and Monte-Carlo shell-model calculations.}


\begin{document}
\section{Introduction}
The strontium ($Z=38$) and zirconium ($Z=40$) isotopes exhibit a sharp shape transition from
a spherical shape at $N=58$ to a strongly prolate-deformed shape at $N=60$. Such a change is
manifested by a decrease of the $2^+$ level energy from 0.8MeV in $^{96}$Sr to 0.15MeV in $^{98}$Sr
and from 1.2MeV in $^{98}$Zr to 0.2MeV $^{100}$Zr. The corresponding $B(E2; 2^+\rightarrow0^+)$
value is raising at the same time about a factor 10 when adding two neutrons only.
The quest of the shape coexistence and quantum phase transition in shape of Zr isotopes
was adressed in many theoretical approaches, including beyond mean-field ones with Gogny and Skyrme forces, 
large-scale shell model (LSSM), Monte-Carlo Shell Model (MCSM)
and algebraic IBM model with configuration mixing, see Ref. \cite{Garcia-Ramos} 
for a recent review.

Such rapid shape transitions are challenging for a theoretical description which has to model
to a great detail the interplay between the stabilizing role of the shell gaps
and the quadrupole correlations tending to deform the nucleus
The shell-model approaches are well suited for an accurate description of such changes, provided a large enough model space
can be handled numerically: it is usually the intruder orbitals coming down 
in the neutron-rich nuclei that are the source of necessary
quadrupole correlations. Among the most successful examples of such spherical-to-deformed modelization
one counts the islands of inversions at $N=20$ and $N=40$, see e.g. \cite{RMP,Utsuno-island,Lenzi2010,PFSDG-U,Miyagi2020}.

In Ref. \cite{Sieja-zr} the shape change in Zr isotopes was adressed in the shell-model calculations
employing a model space outside $^{78}$Ni with proton $1f_{5/2}, 2p_{3/2}, 2p_{1/2}, 1g_{9/2}$
and neutron $1d_{5/2}, 3s_{1/2}, 1g_{7/2}, 2d_{3/2}, 1h_{11/2}$ orbitals. The study permitted a good description
of the spectra and transition rates of odd and even
isotopes up to $N=58$. In particular, it showed a better agreement with experiment as
compared to an earlier work with the $^{88}$Sr core \cite{Holt2000},
which appeared too soft to be used in such calculations in the region \cite{Gumbarcki}.
Nevertheless, the framework of Ref. \cite{Sieja-zr} encountered a difficulty in the description of shape transition from $N=58$
to $N=60$: while the decrease of the $2^+$ energy in $^{100}$Zr was reasonably given, the $E2$ transition rate was
underestimated severly in spite of using a large polarization charge. The problem was then
adressed within the pseudo-SU3 and quasi-SU3 models. The extension of the model space to include
quadrupole partners i.e. the neutron $2f_{7/2}$ and proton $2d_{5/2}$ was suggested to bring the necessary
collectivity. Such calculations could not have been performed in the conventional shell model due to the untractable size
of the configuraton space.
The challenge was later undertaken by the Tokyo group wihin Monte-Carlo shell-model calculations 
permitting to treat extremly large configuration spaces thanks to the application of the Metropolis alghorithm and variational methods
to optimize the single-particle basis vectors \cite{MCSM1, MCSM2}. A beautiful agreement was achieved with experiment in Zr
nuclei, and this as far as $A=110$ \cite{MCSM-zr}, using a large valence space with orbits from two major harmonic oscillator shells
for both neutrons and protons. The authors of \cite{MCSM-zr} also noticed that the abrupt change of the shape in the ground state 
of Zr isotopes can be defined as a quantum phase transition and highlighted the role of the type-II shell evolution in the deformation-driving process.
The same MCSM applied to the Sr isotopes faced however some difficulty, predicting the deformed
configurations to appear at too low neutron number \cite{Regis-Sr}. 

In the present work I revisit the approach of Ref. \cite{Sieja-zr} and apply it to study the Sr isotopes, both
even and odd between the $N=50$ and $N=58$.
Stimulated by the wealth of new data in this region \cite{Regis-Sr, Cruz-PLB, Cruz-N60, Cruz-oddSr, Urban-Sr2021, Clement2016},
the calculations of low-energy spectra, transition rates and spectroscopic factors are performed and discussed.
The approach from Ref. \cite{Sieja-zr} was previously applied in Ref. \cite{Urban-sr} to study $^{92-96}$Sr, with an emphasis
on negative-parity excitations. Recently more shell-model results within this approach 
were presented in Ref. \cite{Urban-Sr2021} but
the calculations were limited to even-even isotopes up to $N=58$. The present work
extends the application to the odd-even Sr isotopes and
in terms of configuration spaces employed.
In addition, the monopole interactions beyond $N=56$
are improved based on spectra of even-odd isotopes in the region. 
This permits to investigate to which extent
the coexistence of spherical and deformed $0^+$ states can be accounted for without the presence of intruders
while having an up-to-date effective interaction. 

The work is organized as follows.
In Section \ref{sm} the shell-model framework and numerical aspects of the calculatons are presented.
Then I discuss the properties of even-odd Sr isotopes in Sec. \ref{odd}. The even-even nuclei are described
in Sec. \ref{even}. Finally,
I address the nature of low-lying $2^+$ states and discuss the possible mechanisms
driving the shape-coexistence and deformation in this region in Sec. \ref{defo}. Conclusions and outlook
are given in Sec. \ref{conc}.

\section{Shell-model framework\label{sm}}
As stated earlier, the calculations are done in the model space outside $^{78}$Ni 
with proton $1f_{5/2}, 2p_{3/2}, 2p_{1/2}, 1g_{9/2}$
and neutron $2d_{5/2}, 3s_{1/2}, 1g_{7/2}, 2d_{3/2}, 1h_{11/2}$ orbitals. The effective interaction
is based on the same Hamiltonian which was used for the study of Zr isotopes in Ref.\cite{Sieja-zr}
and in an earlier study of $^{92-94}$Sr isotopes in Ref. \cite{Urban-sr}, focused 
on the description of low-energy $9^-$ states involving the excitations to the $\nu 1h_{11/2}$ orbital.
This interaction, dubbed $^{78}$Ni-I, was later replaced by the $^{78}$Ni-II version with
a new fit of the proton-proton interaction optimized for the $N=50$ nuclei \cite{Litzinger, Czerwinski-Br}. 
The new fit permitted to reproduce better odd-even neutron-rich nuclei closer to the $^{78}$Ni core. However, 
the proton-neutron and neutron-neutron parts remaining unchanged, the physics of heavier $Z$, even-even nuclei
between $N=50$ and $N=56$ does not vary between the two interactions. One should stress that
the developments of both interactions were focused on nuclei with $N<56$. 
In the present work, the monopole matrix elements $V_{2d_{5/2}-3s_{1/2}}^{T=1}$, $V_{2d_{5/2}-2d_{3/2}}^{T=1}$
$V_{2d_{5/2}-1g_{7/2}}^{T=1}$ are additionally made more attractive
to get a better agreement with experiment also in $^{94,95}$Sr. This modification has little or no effect for 
lighter Sr isotopes where the lowest excited states are dominated by the neutron 
$2d_{5/2}-2d_{5/2}$ interaction.

With two protons less than Zr, the Sr isotopes pose a greater challenge for the shell-model diagonalizations
in the same model space.
In the present work the non-public version of the ANTOINE code developed by E. Caurier is employed,
which permits to treat matrices up to size 10$^{11}$ \cite{Lenzi2010,xenon,PFSDG-U}.
Full space diagonalizations are performed for $^{88-93}$Sr.
For heavier isotopes, up to 8p-8h excitations with respect to the $g_{9/2}$ orbital for protons
and from the neutron $d_{5/2}$ to the rest of the shell are allowed.
A good convergence of spectra of $^{94-96}$Sr is obtained this way.
In $^{96}$Sr 8p-8h truncation gives the matrix size 9.5$\times10^9$ which
requires the most time-consuming calculation performed in this work: computing $3\times0^+$ and  $2\times2^+$
states takes $\sim$188 of CPU hours.

Alternatively, calculations in the seniority scheme are performed for the lowest states of even-even nuclei
to provide more insight into the composition of the wave functions. The $j$-coupled code NATHAN is used for
this purpose. A good convergence of spectra is obtained with seniority 8 in $^{94}$Sr ($j$-coupled dimension 28$\times10^6$) while in $^{96,98}$Sr  
maximally seniority 10 is reached for the $0^+$ states 
leading to $<200$keV convergence of all states.
Converging three $0^+$ states in $^{96}$Sr ($j$-coupled dimension 87.5$\times10^6$)
takes $\sim130$h CPU time. 
The feasibility of the calculations with relatively modest computing resources and CPU time 
is a big advantage of using the present valence space outside the $^{78}$Ni core. 
As will be discussed below, it permits to interpret well the structure 
of Sr nuclei from $N=50$ to $N=58$, before the shape transition takes place 
and the model reaches its application limit.

\section{Results}

\subsection{Properties of odd-even Sr isotopes\label{odd}}
\begin{figure}[H]
\widefigure
\begin{center}
\includegraphics[width=0.35\textwidth]{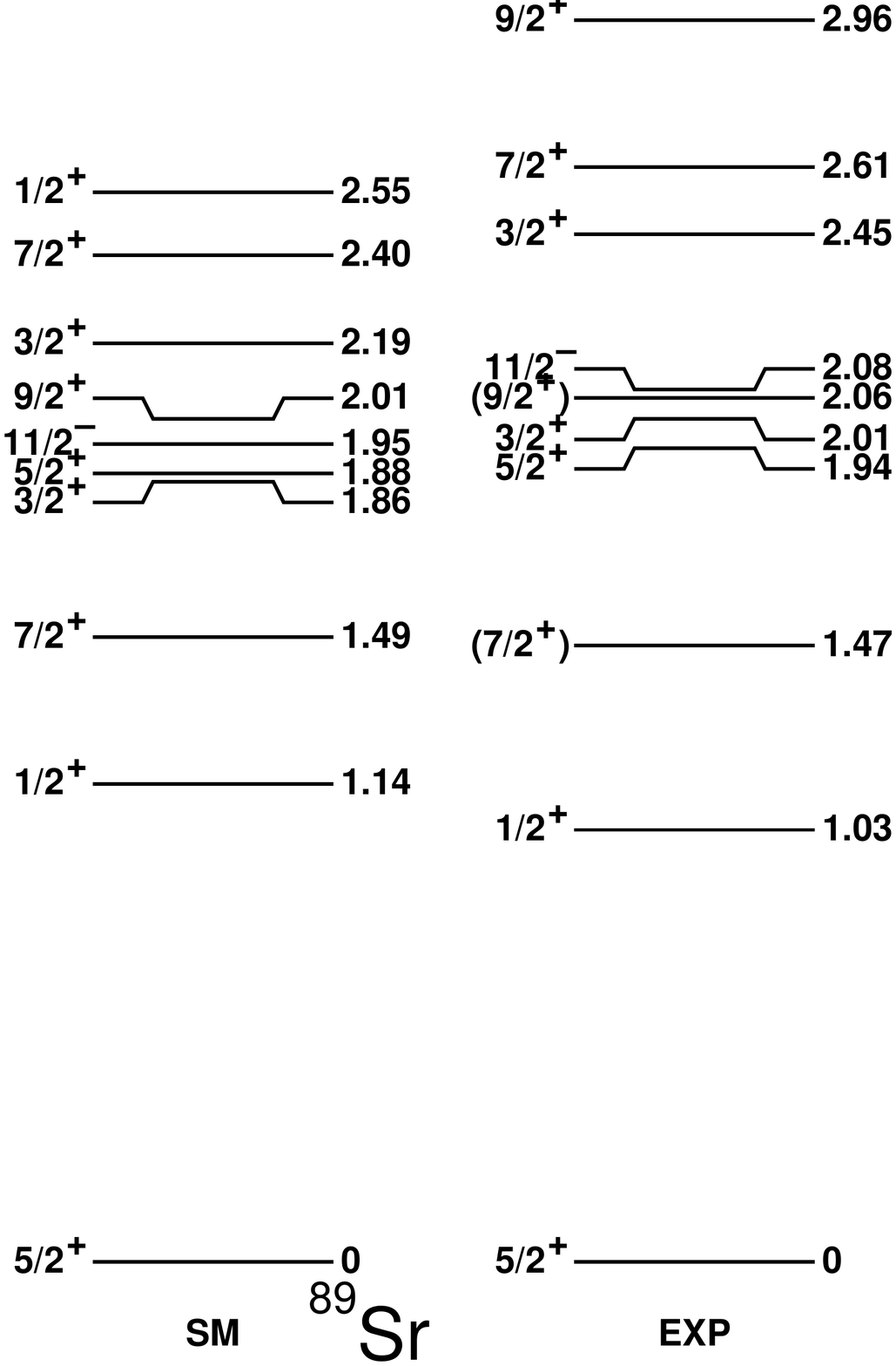}\includegraphics[width=0.35\textwidth]{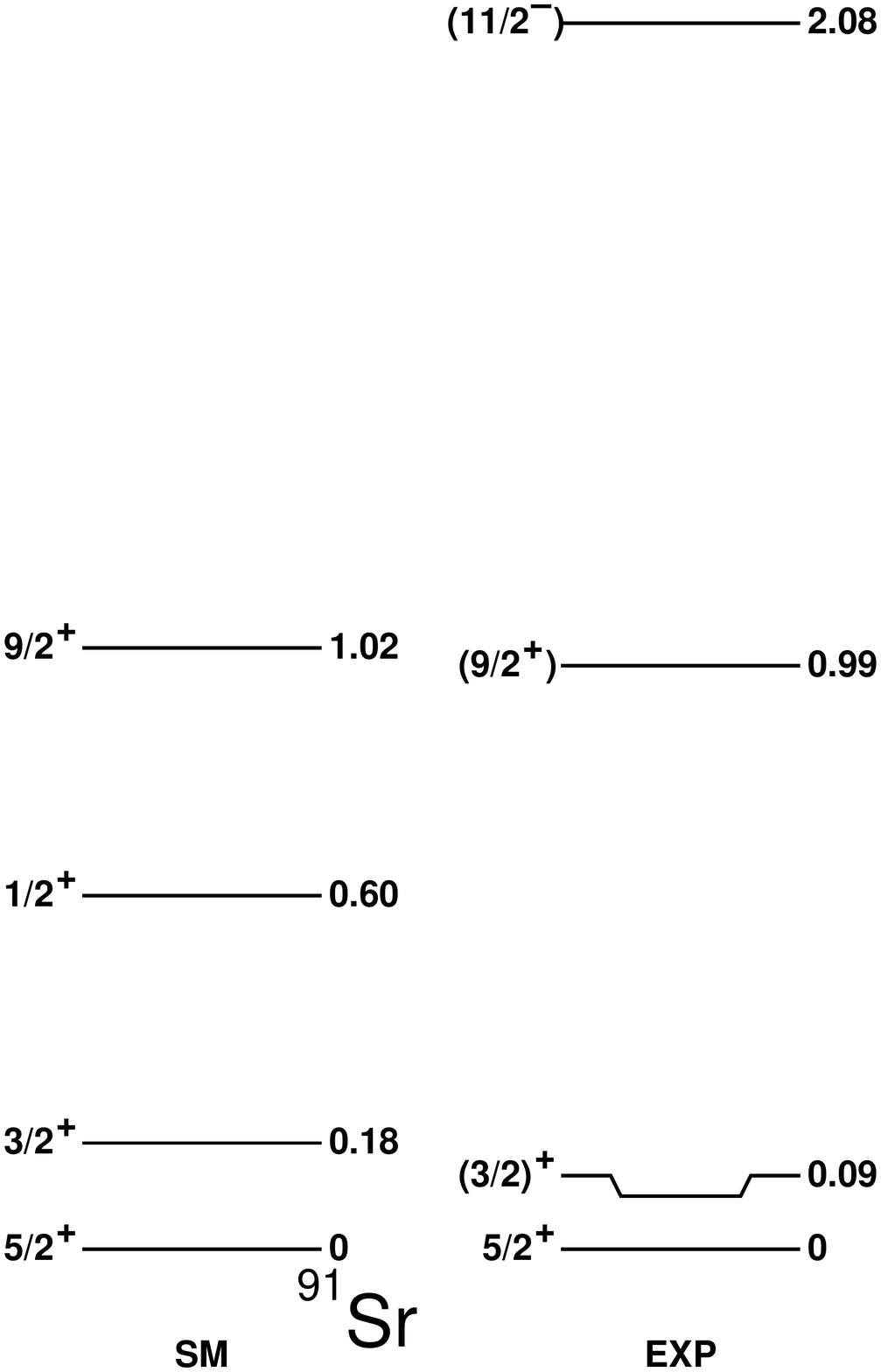}
\caption {Theoretical low-energy spectra of $^{89}$Sr and $^{91}$Sr in comparison to experimental data
from \cite{NNDC}. \label{fig-Sr89}}
\end{center}
\end{figure}

\begin{figure}[H]
\widefigure
\begin{center}
\includegraphics[width=0.35\textwidth]{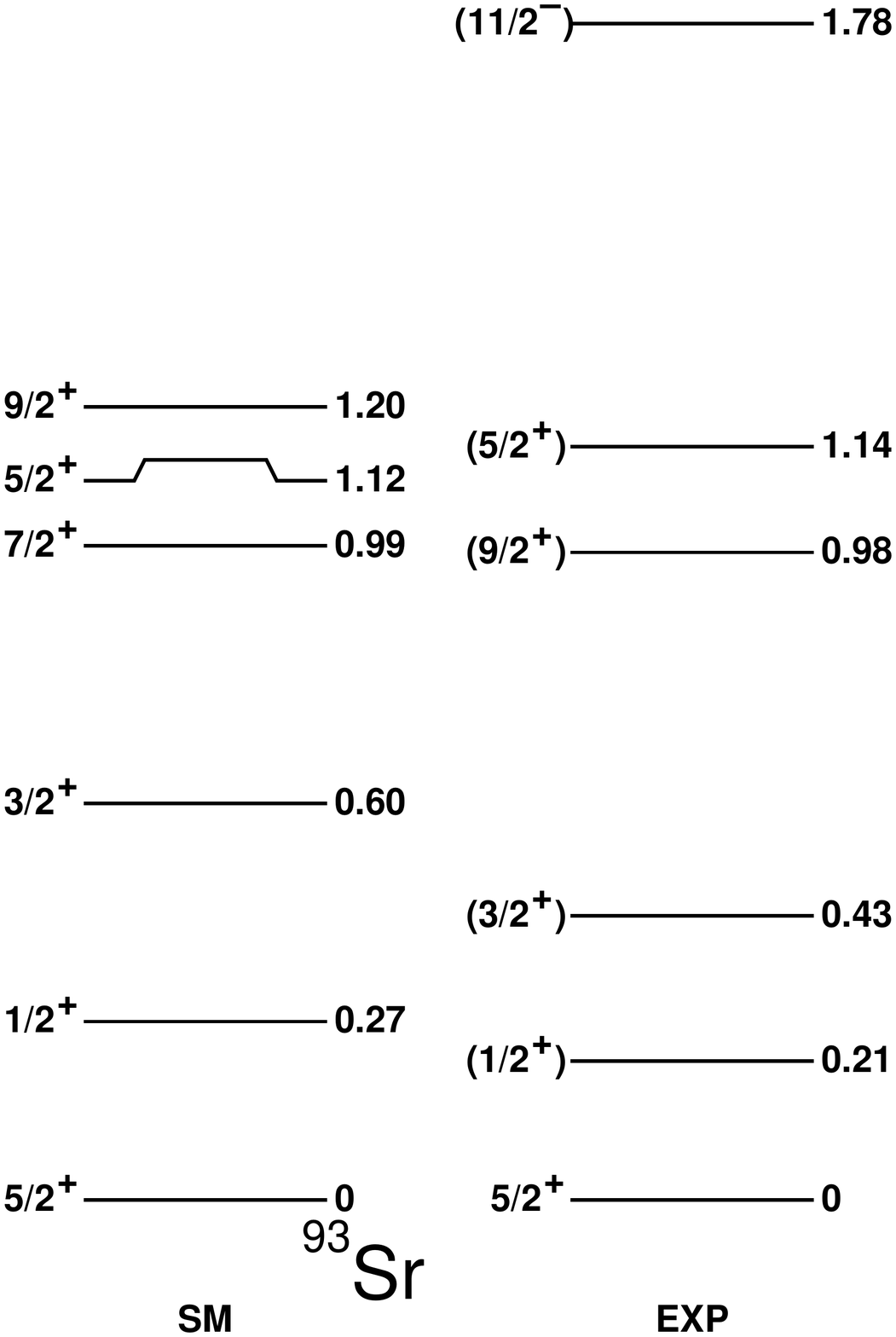}\includegraphics[width=0.35\textwidth]{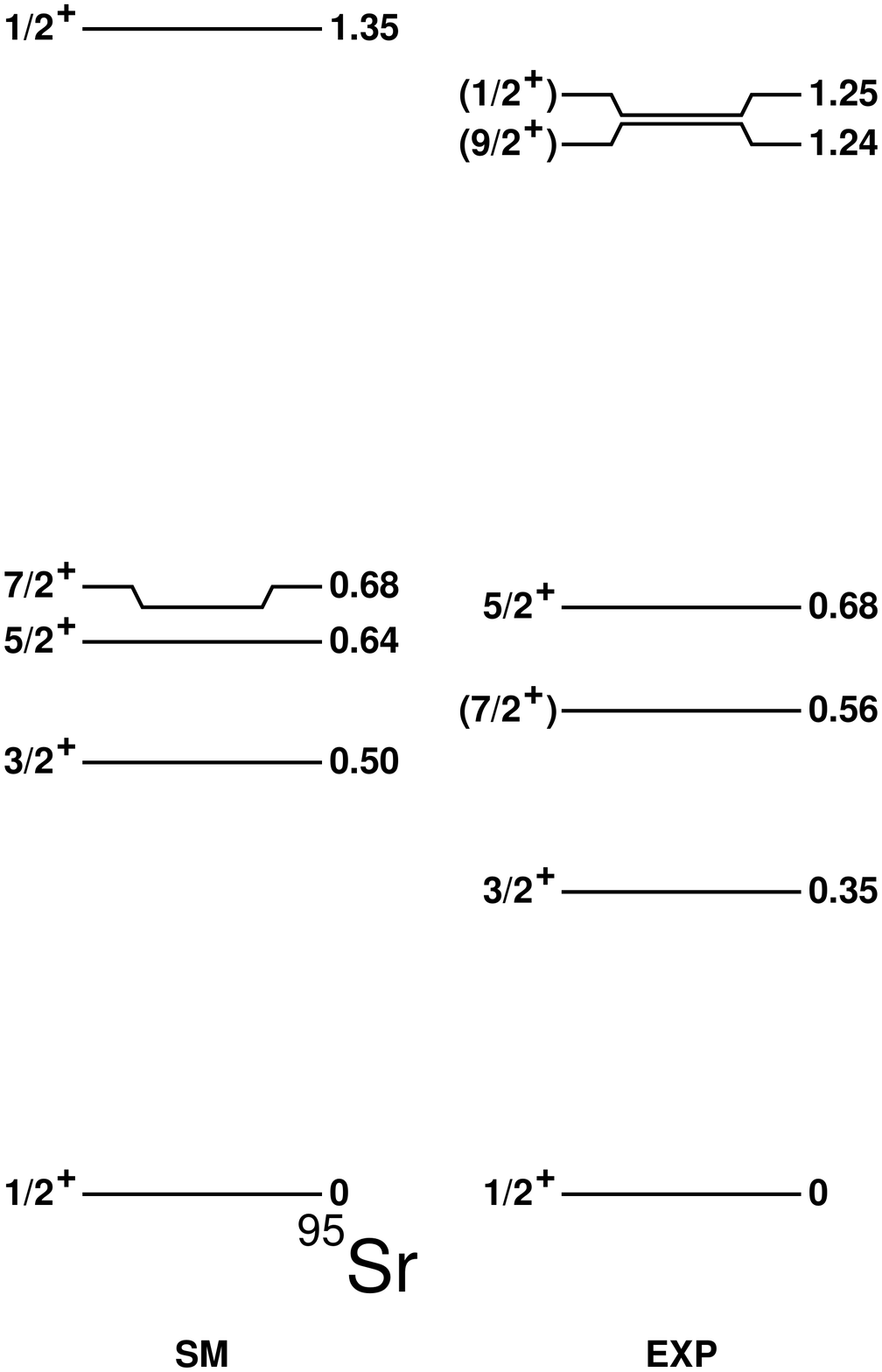}
\caption {Theoretical low-energy spectra of $^{93}$Sr and $^{95}$Sr in comparison to experimental data
from \cite {NNDC, Cruz-oddSr}. \label{fig-Sr93}}
\end{center}
\end{figure}

In Figures \ref{fig-Sr89} and \ref{fig-Sr93} the spectra of odd-even Sr isotopes are presented compared to experimental data.
Those are taken from NNDC \cite{NNDC} for $^{89,91}$Sr while the level schemes established recently
for $^{93,95}$Sr in Ref. \cite{Cruz-oddSr} are plotted in Fig. \ref{fig-Sr93}, supplemented
by the candidates for the characteristic $11/2^-$ excitation taken from \cite{NNDC}. 

Each of the experimentally established
levels finds its counterpart in the shell-model calculations within maximally 250keV.
The rms deviation for the ensemble of levels shown in Figs. \ref{fig-Sr89} and \ref{fig-Sr93}
is of only 140~keV which confirmes the good quality
of the present interaction.
On the other hand, more levels are predicted by theory than assigned experimentally
and various possibilities are not excluded for the spin/parity assignements of a few levels.

To investigate further the single-particle structures in Sr isotopes, the 
spectroscopic factors were computed in $^{93,95}$Sr, to be compared with data obtained in Refs. \cite{Cruz-oddSr, Cruz-N60}.
They are summarized in Table \ref{tab-sf-odd}. For $^{95}$Sr two sets of data are listed (from different reactions) which
correspond to one-neutron removal (first set) and one-neutron addition (second set) spectroscopic factors in the calculations.

\begin{specialtable}[H]
\caption{Spectroscopic factors computed for $^{93,95}$Sr nuclei
compared to experimental data from Table II of Ref. \cite{Cruz-oddSr} (1) and Table IV of Ref. \cite{Cruz-N60} (2).
See text for more details.\label{tab-sf-odd}}
\begin{center}
\begin{tabular}{ccccccc}
\hline
\hline
& experiment &  & & & theory & \\
$J^\pi$ & E(MeV) & $C^2S$& & $J^\pi$ & E (MeV) & $C^2S$\\
\hline
$^{93}$Sr$^{(1)}$ && && &&\\
$5/2^+$    &  0.0     & 3.37(67)& & $5/2^+$ & 0.0 & 3.74\\
$(1/2^+)$  &  0.213 & 0.44(34)& & $1/2^+$ & 0.276 & 0.36\\
$(3/2^+)$  &  0.433 &  -      & & $3/2^+$ & 0.598 & 0.05\\
$(5/2^+)$  &  1.143 & 0.65(15)& & $5/2^+$ & 1.122 & 0.43\\
\hline
$^{95}$Sr$^{(1)}$ && & &&&\\
$1/2^+$    &  0.0    & 0.23(15) & & $1/2^+$ & 0.0   & 0.96\\
$3/2^+$    &  0.352&  -         & & $3/2^+$ & 0.506 & 0.24\\
$5/2^+$    &  0.681& 2.15(50)   & & $5/2^+$ & 0.642 & 3.28\\
$(1/2^+)$  &  1.247& 0.46(15)   & & $1/2^+$ & 1.346 & 0.23\\   
\hline
$^{95}$Sr$^{(2)}$ && & &&&\\
$1/2^+$   &  0.0   & 0.41(9)& & $1/2^+$ & 0.0   & 0.55\\
$3/2^+$   & 0.352& 0.53(8)  & & $3/2^+$ & 0.506 & 0.65\\
$5/2^+$   & 0.681& 0.16(3)  & & $5/2^+$ & 0.642 & 0.07\\
\hline
\hline
\end{tabular}
\end{center}
\end{specialtable}

As can be seen, the magnitude of the spectrosopic factors is fairly reproduced. The values are close to
experiment in the case of $^{93}$Sr and for one-neutron addition data in $^{95}$Sr.
The largest discrepancy is found for $1/2^+$ and $5/2^+$ states in $^{95}$Sr, in the case of one-neutron removal. 
As the computed spectroscopic factors are too large, the occupation of these orbits in the wave-function of the 
$0^+$ ground-state of $^{96}$Sr 
are probably too large, too, and higher orbitals should be better populated.
The structure of the $0^+$ states of Sr isotopes is discussed in the next Section.

\begin{specialtable}[H]
\caption{Occupation numbers and magnetic moments in selected low-energy states in odd-Sr isotopes.
The occupations lower than 0.01 particle are rounded to zero.\label{tab-wf-odd}}
\scriptsize
\begin{tabular}{cccccccccccccc}
\hline
\hline
$N$ & $J^\pi$ & $E$ (MeV) & $\mu (\mu_N)$ &$\mu_{exp}(\mu_N)$ & $\pi f_{5/2}$ & $p_{3/2}$ & $p_{1/2}$ & $g_{9/2}$ & $\nu d_{5/2}$ & $s_{1/2}$ & $g_{7/2}$ & $2d_{3/2}$ & $h_{11/2}$ \\
\hline
51 & $1/2^+$  &  1.14 & -1.32 & -         & 5.48 & 3.53 & 0.42 & 0.57 & 0.08 & 0.91 & 0.0 & 0.0 & 0.01\\
   & $3/2^+$  &  1.86 &  0.22 & -         & 5.30 & 3.33 & 0.85 & 0.52 & 0.66 & 0.0  & 0.0 & 0.32& 0.01\\
   & $5/2^+$  &  0.0  & -1.19 & -1.1481(8)& 5.59 & 3.30 & 0.49 & 0.61 & 0.97 & 0.0  & 0.0 & 0.0 & 0.02\\
   & $7/2^+$  &  1.49 &  0.15 & -         & 5.27 & 3.19 & 1.10 & 0.44 & 0.97 & 0.0  & 0.0 & 0.0 & 0.02\\
   & $9/2^+$  &  2.01 &  0.47 & -         & 5.06 & 3.35 & 1.15 & 0.44 & 0.98 & 0.0  & 0.0 & 0.0 & 0.01\\
   & $11/2^-$ &  1.95 &  1.55 & -         & 5.68 & 2.85 & 0.29 & 1.16 & 0.69 & 0.0  & 0.0 & 0.0 & 0.29\\
\hline
53 & $1/2^+$  &  0.60 & -0.99 & -         & 5.37 & 3.27 & 0.77 & 0.59 & 1.88 & 0.81 & 0.08 & 0.11 & 0.10\\ 
   & $3/2^+$  &  0.18 & -0.48 & -0.347(17)& 5.44 & 3.13 & 0.79 & 0.64 & 2.57 & 0.20 & 0.05 & 0.11 & 0.07\\  
   & $5/2^+$  &  0.0  & -0.94 & -0.885(2) & 5.44 & 3.15 & 0.75 & 0.65 & 2.60 & 0.09 & 0.08 & 0.11 & 0.12\\
   & $7/2^+$  &  1.28 &  0.44 & -         & 5.22 & 3.16 & 1.14 & 0.48 & 2.61 & 0.10 & 0.07 & 0.12 & 0.10\\
   & $9/2^+$  &  1.01 & -1.01 & -         & 5.40 & 3.17 & 0.82 & 0.61 & 2.71 & 0.07 & 0.04 & 0.10 & 0.08\\ 
   & $11/2^-$ &  2.15 &  1.18 & -         & 5.54 & 2.89 & 0.52 & 1.05 & 2.14 & 0.09 & 0.10 & 0.13 & 0.53\\
\hline
55 & $1/2^+$  &  0.27 & -0.96 &(-1.02(6)) & 5.01 & 3.31 & 1.06 & 0.62 & 3.51 & 0.83 & 0.21 & 0.28 & 0.17\\ 
   & $3/2^+$  &  0.60 &  0.09 & -         & 5.08 & 3.28 & 1.02 & 0.62 & 3.41 & 0.97 & 0.20 & 0.27 & 0.14\\ 
   & $5/2^+$  &  0.0  & -0.88 &-0.7926(12)& 5.18 & 3.15 & 1.02 & 0.65 & 4.04 & 0.27 & 0.20 & 0.29 & 0.19\\ 
   & $7/2^+$  &  0.99 & -0.33 & -         & 5.06 & 3.31 & 1.02 & 0.61 & 3.47 & 0.84 & 0.21 & 0.34 & 0.14\\
   & $9/2^+$  &  1.20 & -0.32 & -         & 5.09 & 3.23 & 1.16 & 0.52 & 3.96 & 0.35 & 0.18 & 0.33 & 0.17\\
   & $11/2^-$ &  1.95 &  1.02 & -         & 5.28 & 2.96 & 0.78 & 0.96 & 3.43 & 0.26 & 0.24 & 0.35 & 0.72\\
\hline
57 & $1/2^+$  &  0.0  &-0.58& -        & 4.57 & 3.49 & 1.42 & 0.52 & 5.02 & 0.98 & 0.34 & 0.45 & 0.20\\
   & $3/2^+$  &  0.51 & 0.49& -        & 4.95 & 3.32 & 1.15 & 0.57 & 4.73 & 0.67 & 0.33 & 1.06 & 0.21\\  
   & $5/2^+$  &  0.64 &-0.46& -0.537(2)& 4.45 & 3.60 & 1.48 & 0.47 & 4.53 & 1.52 & 0.33 & 0.41 & 0.21\\
   & $7/2^+$  &  0.68 & 0.73& -        & 4.80 & 3.30 & 1.23 & 0.66 & 4.64 & 0.64 & 1.09 & 0.42 & 0.21\\ 
   & $9/2^+$  &  1.47 & 0.93& -        & 4.78 & 3.37 & 1.22 & 0.63 & 4.35 & 0.92 & 1.06 & 0.50 & 0.17\\
   & $11/2^-$ &  2.46 & 0.03& -        & 4.97 & 3.22 & 1.10 & 0.71 & 4.64 & 0.45 & 0.40 & 0.49 & 1.02\\ 
\hline
\hline
\end{tabular}
\end{specialtable}

In Table \ref{tab-wf-odd}
the occupations of the proton and neutron orbitals are listed for the low-energy excitations with $1/2^+-9/2^+$ and
$11/2^-$ spin/parity. In addition to the spectroscopic factors listed above, the magnetic moments were computed
for all low-energy states (using 0.7 quenching on spin part of the $M1$ operator). 
They are listed along with the occupation numbers and compared to experimental values
from \cite{NNDC} when possible. 
The agreement between theoretical and experimental magnetic moments is fairly satisfying.
Note that the 213keV level in $^{93}$Sr is assigned as $(9/2)^+$ in NNDC while it was suggested to be $(1/2^+)$
in Ref. \cite{Cruz-oddSr} which agrees better with shell-model predictions. Also the value of the magnetic moment,
displayed in the Table in parantheses, corresponds well to that of the computed $1/2^+$ level, supporting further 
the spin/parity assignement from \cite{Cruz-oddSr}.

As can be noted, the neutron occupations
reveal a clearly single-particle structure for the ground state and the first $1/2^+$ state in $^{89}$Sr. 
The occupation of the $2d_{3/2}$ in the $3/2^+$ and of the $1h_{11/2}$ in the $11/2^-$ is of the order of $30\%$
thus those states are mostly resulting from coupling of the odd neutron to proton excited states.
The first $7/2^+$ state, with no particle in $1g_{7/2}$, is predicted due to a coupling of the proton $2^+$
with the $2d_{5/2}$ neutron with $95\%$ probability. The occupation of the $1g_{7/2}$ orbital grows to 0.7 particle
in the second excited $7/2^+$ (not shown in the Table) predicted at 2.4MeV. The energy of this orbital was 
estimated to be around 2MeV in the $^{78}$Ni core and its evolution with the neutron number is crucial for the development
of the collectivity, as will be outlined below. Unfortunately, no experimental information on its position is currently available
in the region from the experimentally available spectroscopic factors.
With the increasing neutron number one observes an increase of collectivity of the lowest states, manifested 
by more spread occupancies on both proton and neutron sides. Still, up to $N=55$, the lowest excitations are
based on neutrons in $2d_{5/2}-3s_{1/2}$ orbitals. The mixing of neutrons from $2d_{5/2}$ and $3s_{1/2}$ shells
reflects the lack of a shell closure at $N=56$ in Sr isotopes which can be inferred from the $2^+$ systematics.
The occupation of the $1h_{11/2}$ orbital
in the $11/2^-$ level increases steadily from 0.3 to 1.0 particle at $N=57$. However, no experimental
information on the position of the $11/2^-$ excitation is available in this nucleus to confirm the predicted
tendency. After passing the $N=56$, also the first excited states $3/2^-$ and $7/2^-$ 
have around 1 particle in the $2d_{3/2}$ and $1g_{7/2}$ orbitals, respectively. Nevertheless, their wave functions 
remain spread over many components with probabilities less than 10\%.
    
It is worth mentionning that similar shell-model calculations were carried out in Refs. \cite{Cruz-oddSr, Cruz-N60, Cruz-PLB}
exploring smaller configuration spaces. The present results for $^{93}$Sr and $^{95}$Sr
seem more satisfactory in their prediction of the position of the first excited 
$9/2^+$ state (see Figs. 13 and 16 of Ref. \cite{Cruz-oddSr}). This state contains a considerable
admixture of the proton $g_{9/2}$ orbital, crucial for a proper description
of single-particle and collective excitations in this region of nuclei. Its exclusion
from the model space used in \cite{Cruz-oddSr} may be thus responsible for a too high position of
the excited $9/2^+$ states in $^{93,95}$Sr.

\subsection{Low-energy spectra of even-even Sr isotopes\label{even}}
\begin{figure}[H]
\begin{center}
\includegraphics[width=0.35\textwidth]{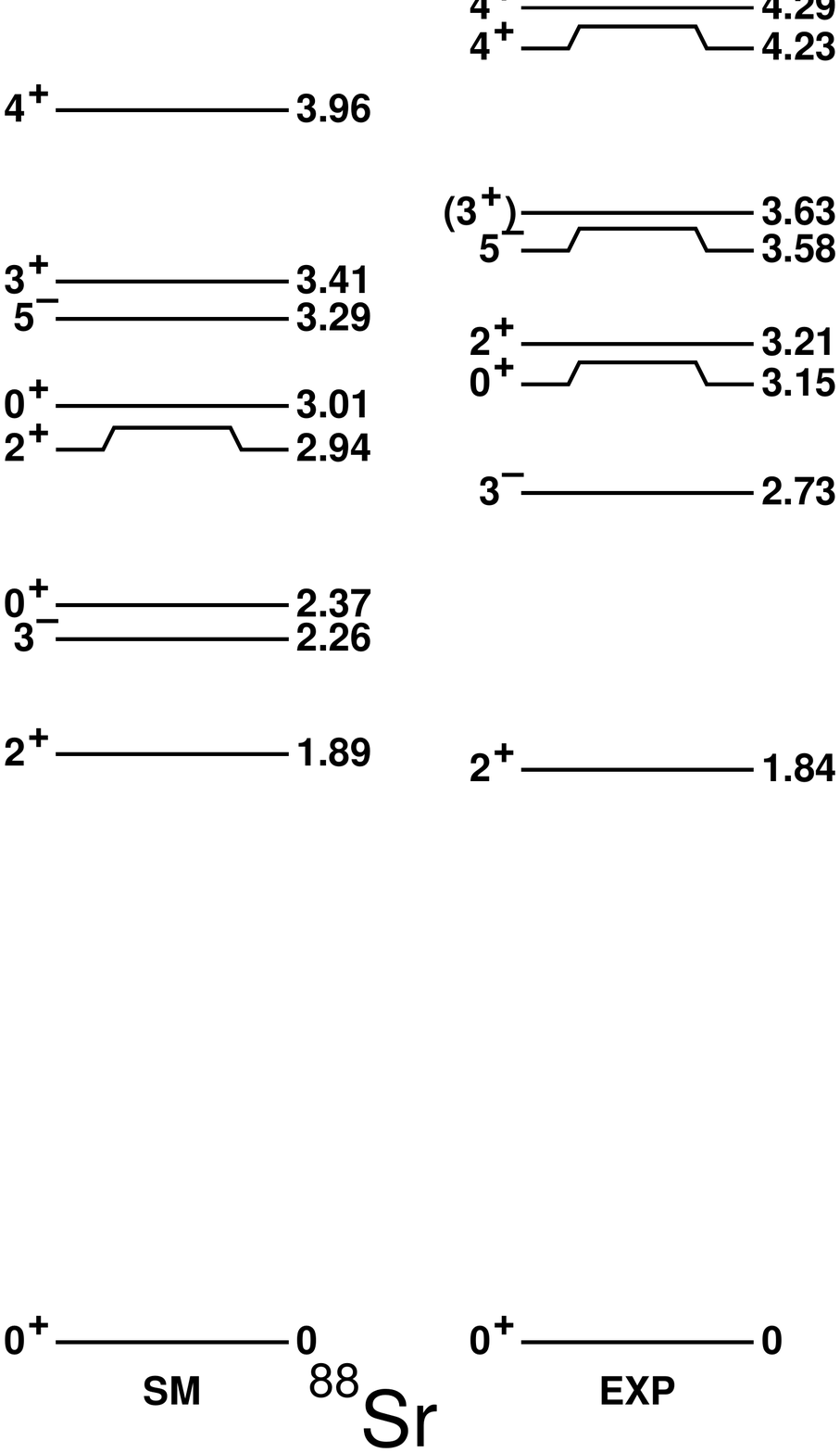}\includegraphics[width=0.35\textwidth]{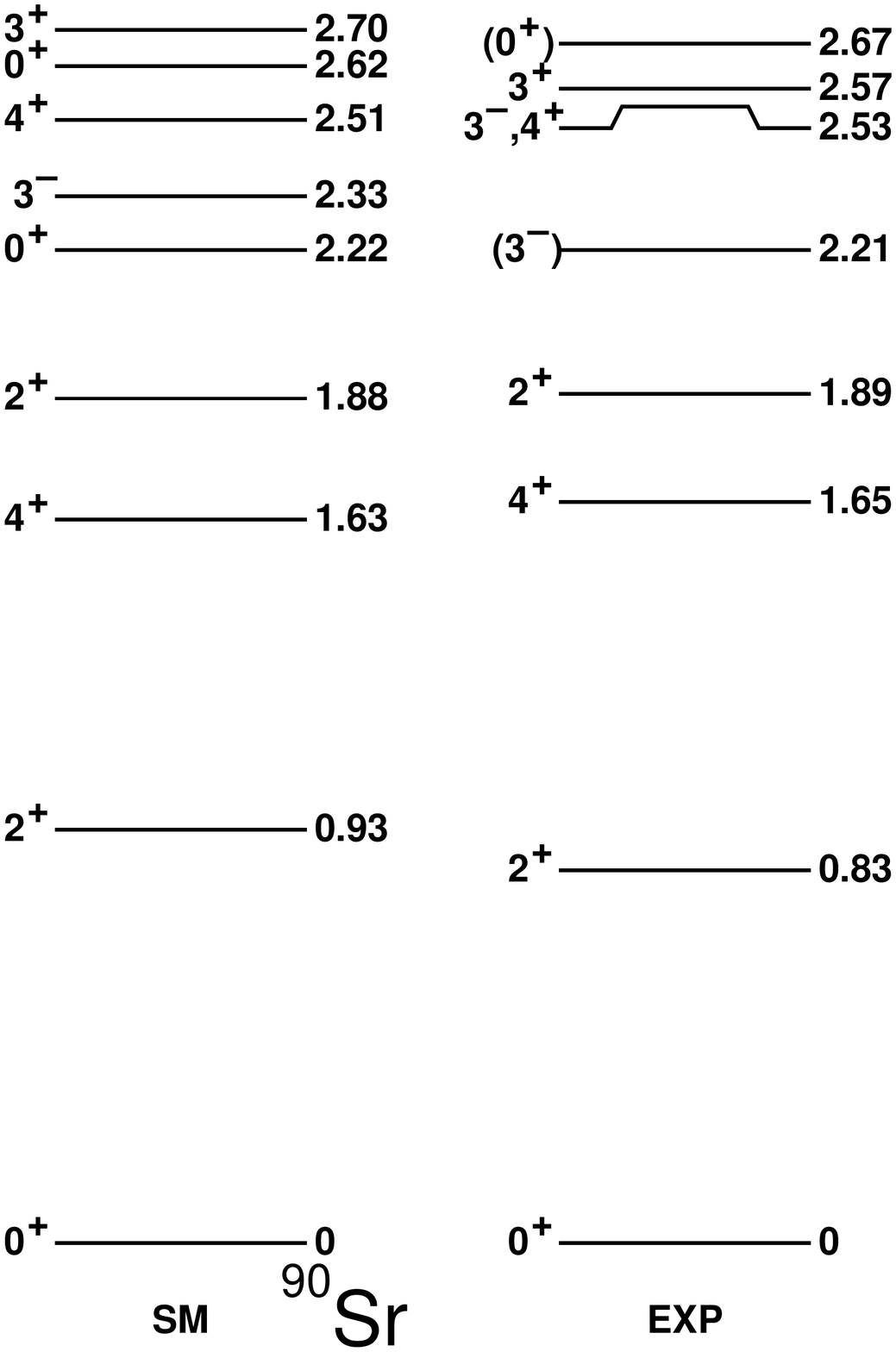}
\caption {Theoretical low-energy spectra of $^{88}$Sr and $^{90}$Sr in comparison to experimental data
from \cite{NNDC}. \label{fig-Sr88}}
\end{center}
\end{figure}

\begin{figure}[H]
\begin{center}
\includegraphics[width=0.35\textwidth]{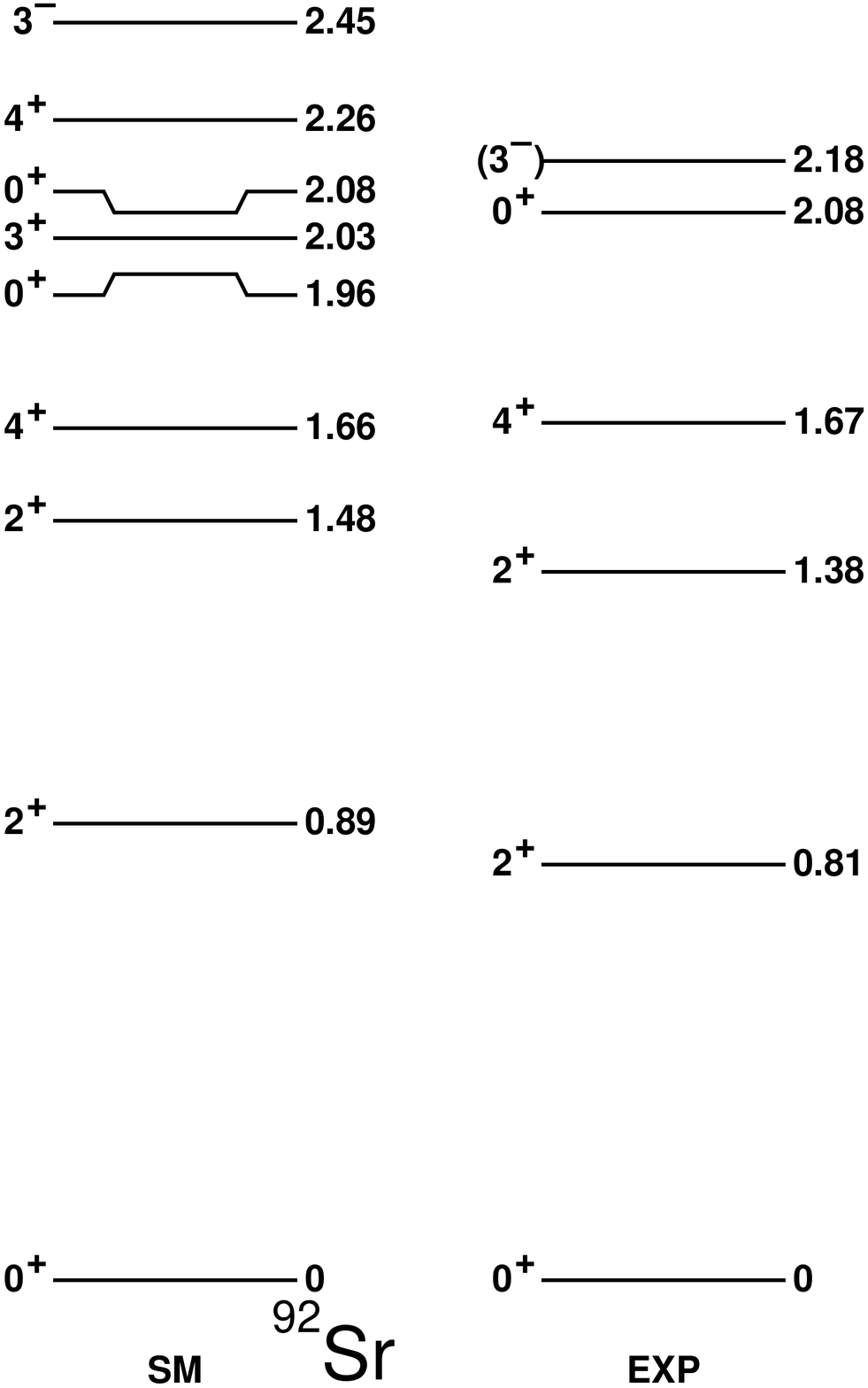}\includegraphics[width=0.35\textwidth]{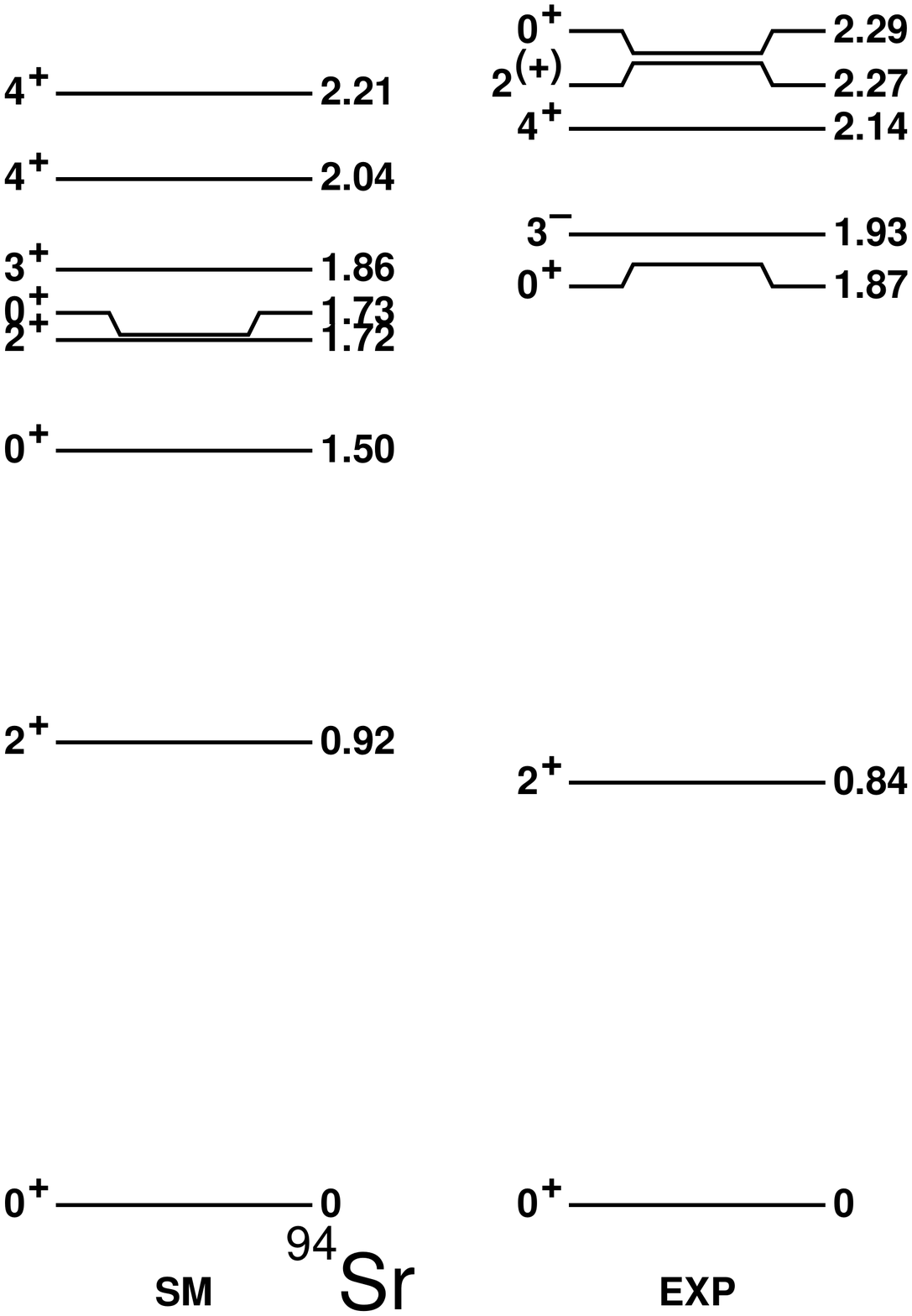}
\caption {The same as in Fig. \ref{fig-Sr88} but for $^{92}$Sr and $^{94}$Sr.\label{fig-Sr92}}
\end{center}
\end{figure}

In Figures \ref{fig-Sr88} and \ref{fig-Sr92} the low-energy spectra of even-even Sr between $N=50$ and $N=56$
are shown.
The overall agreement with experiment is very good in lighter Sr, with larger uncertainties at subshell closures. 
Nonetheless, the rms deviation for the levels shown is 300keV, twice larger than for odd nuclei. 
In $^{88}$Sr the shell model predicts the $3^-$ excitation 470keV lower 
than the experiment, with $70\%$ of $\pi 2f_{5/2}^6 2p_{3/2}^3 1g_{9/2}^1$ configuration. 
The  $5^-$ excitation is based on the same configuration ($74\%$)
but predicted closer to the experiment (within 290keV).  
In addition, the $9/2^+$
excited level in the neighbouring $^{87}$Rb ($Z$=37) is predicted at 1366keV to be compared to the experimental 
value of 1577.9keV. 
The larger disagreement for the $3^-$ level should thus not be related
to the position of the $1g_{9/2}$ orbital or the monopole part of the interaction involving it.  

Interestingly, the $3^-$ state in $^{90}$Sr fits much better the experimental candidates,
whether the first $3^-$ is the 2.21MeV or 2.53MeV level. Here the $\pi 2f_{5/2}^6 2p_{3/2}^3 1g_{9/2}^1$
component counts for only $40\%$ and one notes the neutron $2d_{5/2}^1-1h_{11/2}^1$ configuration contributes to the
wave function. The occupation of the $1h_{11/2}$ grows from 0.21 at $N=52$ to 0.48 at $N=56$ in the $3^-$ states.
The energy of this excitation deviates more and more with increasing neutron number, suggesting
octupole collectivity starts playing a role in heavier Sr isotopes, as it is the case in heavier-$Z$ nuclei.

\begin{figure}[H]
\begin{center}
\includegraphics[width=0.4\textwidth]{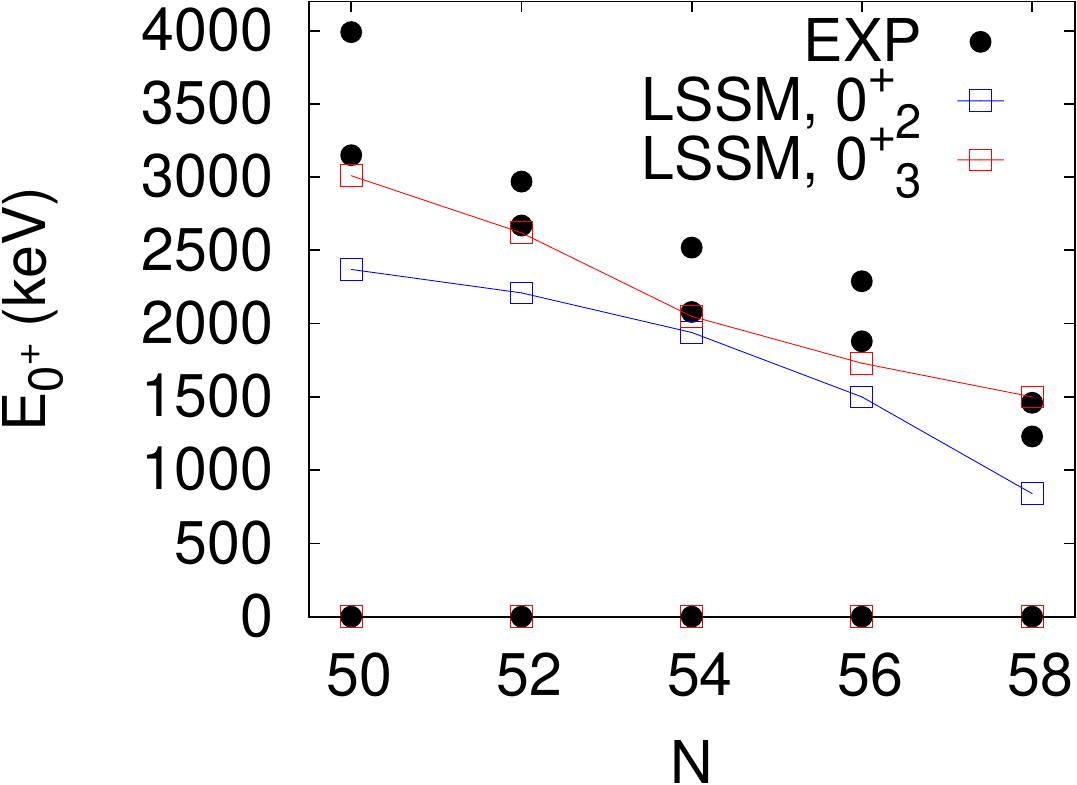}
\caption{Systematics of the $0^+$ states in Sr isotopes. Experimentally known levels are indicated in black while shell-model
results in red and blue. Lines are added to guide the eye. \label{fig-0}}
\end{center}
\end{figure}

Another discrepancy in $^{88}$Sr concerns the first excited $0^+$, predicted well below the first known state, while the 2nd
shell-model $0^+$ fits very well the experiment. This situation propagates along the chain, see Fig. \ref{fig-0}.
The systematics of the $0^+$ states was also discussed in the previous work (see Fig. 11 of Ref. \cite{Urban-Sr2021})
and a similar disagreement with experiment was observed. 
Present calculations are performed with an improved shell-model interaction, 
as described in Sec. \ref{sm}, and provide an overall better agreement with
experiment in heavier Sr compared to calculations from Ref. \cite{Urban-Sr2021}. 
The problem of low-lying $0^+$ states remains anyway. As can be taken from Fig. \ref{fig-0},
it is the 2nd excited $0^+$ from theory that follows closely the experimental data while
the first theoretical $0^+$ seems to have no counterpart. 

Based on data from \cite{Urban-Sr2021}
it was not possible to clarify whether a low-energy $0^+$ excitation is systematically not observed in experiment
or the shell model underestimates the energy of the first excited $0^+$ state.
To get more insight into this issue, the distribution of one-particle removal spectroscopic factors was computed for
the low-spin states in $^{94}$Sr.  
The values for the lowest states are shown in Table \ref{tab-Sr94sf} along with the data from \cite{Cruz-oddSr}.
   
\begin{specialtable}[H]
\caption{Spectroscopic factors in $^{94}$Sr, see text for more details.\label{tab-Sr94sf}}
\begin{center}
\begin{tabular}{cccccc}
\hline
\hline
& experiment & & &theory & \\
$J^\pi$ & E(MeV) & $C^2S$ & $J^\pi$ & E(MeV) & $C^2S$\\
\hline\\
$0^+_1$   & 0.0    & 0.336(7) & $0^+_1$ & 0.0 & 0.55\\
$0^+_2$   & 1.88   & 0.067(4) & $0^+_2$ & 1.50  & 0.07 \\
$0^+_3$   & 2.29   & 0.105(6) & $0^+_3$ & 1.73  & 0.11 \\
$2^+_1$   & 0.84   & 0.725(25)& $2^+_1$ & 0.92  & 1.04 \\
$3^+_1$   & 2.41   & 0.99(3)  & $3^+_1$ & 1.86  & 1.88 \\ 
\hline
\hline 
\end{tabular}
\end{center}
\end{specialtable}

The agreement between the computed and experimental spectroscopic factors is fair for the $0^+$ states: 
though the value of the ground state is slightly overshot, 
the differences in the magnitude between the 3 states are particularily well reproduced. 
One can thus conclude that the first three LSSM states correspond indeed to their experimental partners. This is true at least 
in $^{94}$Sr as no such spectroscopic factors are available for $N\le 56$.
The model underestimates excitation energies which can be due to inaccuracies of diagonal and non-diagonal 
matrix elements of pairing interactions. Such a problem was avoided 
in Zr isotopes where the $2p_{1/2}, 1g_{9/2}$ orbitals are mostly involved in the $0^+$s, 
contrary to Sr with large $1f_{5/2}-2p_{3/2}$ mixing in the wave-functions.   
I note, in passing, in recent MCSM calculations in $^{94}$Sr nucleus
a triaxially deformed band predicted on the $0^+_2$ state was located below 1.5MeV. Thus at a similar energy as the first excited $0^+$
predicted in the present calculations.

\begin{figure}[H]
\begin{center}
\includegraphics[width=0.4\textwidth]{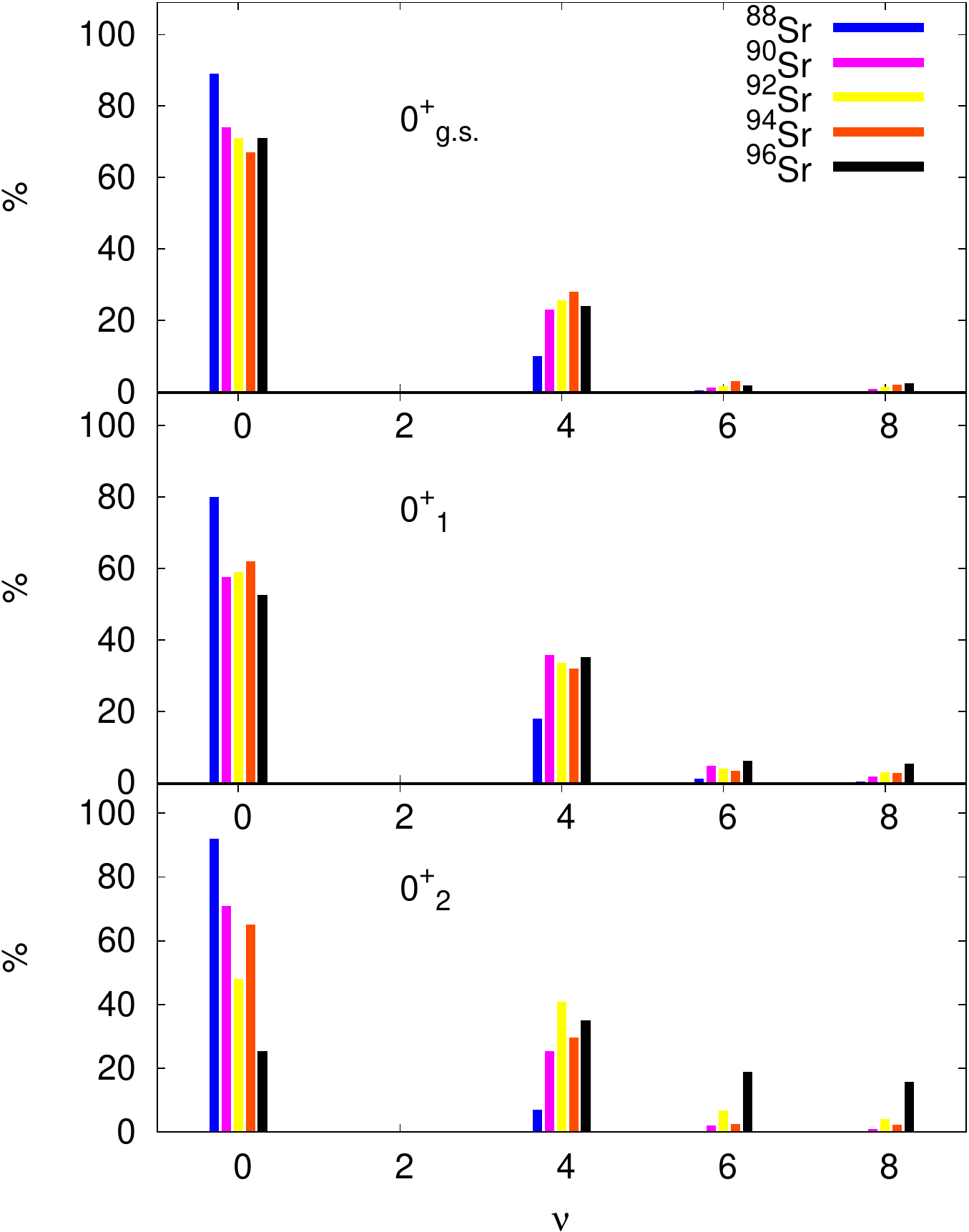}
\caption{Seniority composition of the $0^+$ states in Sr isotopes.\label{0-sen}}
\end{center}
\end{figure}

The lowest $0^+$ were also computed
using the $j$-coupled code with seniority truncation which is the most efficient scheme to converge
multiple $0^+$ excitations in spite of large sizes of matrices. The composition
of the $0^+$ states is shown in Fig. \ref{0-sen}. The seniority zero component decreases with mass, still is
predicted to dominate in all $0^+$ states and in all isotopes between $N=50$ and $N=56$. 
Seniority 4 does not exceed $40\%$ while $\nu=6,8$ are minor. 
Seniority $\nu=10$ and higher components are negligible 
(and not shown in the Figure), which explains the fast convergence of the
computed states in terms of number of broken nucleonic pairs.
At $N=58$ the higher seniority components are also minor for the first two $0^+$ states
but the 3rd $0^+$ changes its structure: it is dominated by $\nu=4$ and $\nu=6,8$ reach $20\%$.
This change of structure is consistent with the 
prediction that one of the excited $0^+$ states should be deformed at $N=58$.

As for the dominating configurations, $0^+$ states in $^{88}$Sr are composed of: \\
$0^+_{g.s.}$: $60\%$ of $1f_{5/2}^62p_{3/2}^4$,\\
$0^+_1$: $32\%$ of $1f_{5/2}^42p_{3/2}^21g_{9/2}^2$,\\
$0^+_2$: $55\%$ of $1f_{5/2}^42p_{3/2}^22p_{1/2}^2$. Adding two neutrons in $^{90}$Sr
one finds the same proton components coupled to $2d_{5/2}^2$ neutrons,
in the same order but with a smaller percentage. The situation changes 
in $^{92}$Sr, where an exchange of major configurations takes place: the first excited $0^+$
is now dominated by the $1f_{5/2}^42p_{3/2}^22p_{1/2}^2$ component coupled to $2d_{5/2}^4$ neutrons,
while the second $0^+$ contains more of $1f_{5/2}^42p_{3/2}^21g_{9/2}^2$ configuration. 
The percentages of various components continue to drop with neutron number till 
reaching less than $10\%$ for the dominating components in the ground state of $^{94}$Sr. 
All the $0^+$ states computed in this nucleus
are based on the $1f_{5/2}^42p_{3/2}^22p_{1/2}^2$ configuration with 9$\%$, 27$\%$ and 28$\%$ 
respectively and neutrons occupy mostly $1d_{5/2}$ and $3s_{1/2}$ orbitals. It is interesting to notice
that the ground state is the purest in terms of seniority 
but the most fragmented over different configurations.  
At $N=58$ the same structures in the wave-functions can be found in the first two $0^+$ states with 
$18\%$ and $23\%$. In the 3rd $0^+$ none of the configurations is privileged. Different particle-hole 
components do not exceed $5\%$ confirming this state is the most collective of all the calculated $0^+$s.
At the same time the occupation of the $1g_{9/2}$ and $1g_{7/2}$ orbitals doubles in this state
with respect to the ground state, reaching 1.43 and 1.41 particle, respectively.
A much smaller number than the occupation predicted in the deformed $0^+$ state in $^{98}$Zr within the MCSM, where
the $1g_{9/2}$ has more than 3 protons. Moving to $N=60$ these occupations do not grow in the ground state 
which is also dominated by $\nu=0$ component, contrary to the expectations.
This problem is discussed in the next section devoted to the shape change
at $N=60$.
 
Table \ref{tab-Sr94sf} reports as well the values of spectroscopic factors for higher spin states compared to data in $^{94}$Sr.
The spectroscopic factor calculated here for the $2^+$ state is consistent with exprimental value
and the excitation energy of the state agrees very well. 
The largest discrepancy among the computed states concerns the first $3^+$ level: here the theoretical energy is  
550keV too low and the spectroscopic factor twice too large; the LSSM calculations reported in 
Ref. \cite{Cruz-oddSr} agreed better in energy but overestimated the spectroscopic factor
nearly a factor 3. Present calculations predict other excited $3^+$ states 
in the vicinity of the experimental value with much lower spectroscopic factors.
Unfortunately, experimentally no more unambigous spin assignements are available for the $3^+$ states
to compare the total strength and its distribution.    
As discussed before in Ref. \cite{Urban-Sr2021} along the Sr chain, the lowest $3^+$ energy is predicted by LSSM in $^{92}$Sr.
It is also in this nucleus where the $3^+$ state bears some characteristics of the non-axially deformed state.
On the contrary, at $N=56$, the $B(E2)$ transitions from the $3^+$ state to the $2^+_2$ are not particularily  
strong. 
No $3^+$ excitation was reported from MCSM in Ref. \cite{Regis-Sr} but one can expect a low-energy $3^+$ state
along with the prediction of a triaxial band. It would be of high interest to investigate 
further wether triaxiality is present or not at $N=56$ 
and wether it could explain the magnitude of the spectroscopic factor of the $3^+$ extracted from experiment.

\subsection{Collectivity in the Sr chain towards the $N=60$ \label{defo}}
The early shell-model work of Federman and Pittel \cite{Federman79}, in a small configuration space, pointed out the rapid decrease
of the $2^+$ energy around $N=60$ in connection to the important role of the strong, attractive $\pi 1g_{9/2}-\nu 1g_{7/2}$
interaction (the so-called Spin-Orbit Partners (SOP) mechanism). However,
this mechanism appeared insufficient to explain the deformation in Zr isotopes in Ref. \cite{Sieja-zr}
so it was concluded orbitals from adjacent shells play a significant role in shaping the nuclei of this region.
The suggestion was to add to the model space the quadrupole-driving orbitals $\nu 2f_{7/2}$ and
$\pi 2d_{5/2}$ to create quasi-SU3 blocks for neutrons and protons operating in addition to two pseudo-SU3
blocks formed by the lower shells. Addition of only one of the two orbitals was shown to be possibly sufficient to
reproduce the enhancement of the $B(E2)$ value observed between $^{98}$Zr and $^{100}$Zr.
Alternatively, it was debated if the increased population of the $\nu 1g_{7/2}$ at $N=60$ can be due to the 
promotion of the particles from the extruder neutron $\nu 1g_{9/2}$ orbital (across the $N=50$ gap), see
\cite{Urban-zr100} and references therein.

In fact, the MCSM calculations presented in Ref. \cite{MCSM-zr} can provide an answer to these questions. The model space used
contained both $\nu 1g_{9/2}$ and $\nu 2f_{7/2},3p_{3/2}$ as well as $\pi 2d_{5/2}$ orbitals.
The authors highlighted the role of the type-II shell evolution 
which confirms the original idea of Federman and Pittel on the important role of the SOP mechanism with the $\pi g_{9/2}$
and $\nu g_{7/2}$ attraction as the primary reason for the shape change. 
While it was not stated explicitely from which orbitals the additional quadrupole correlations came from, the 
inspection of the effective single-particle energies (ESPE) and occupation numbers
dispalyed in Fig. 3 of the same work provides the necessary
insight. First, the proton $1d_{5/2}$ occupation is non-zero in the deformed states. As was shown 
in the island of inversion study in Ref. \cite{Lenzi2010}, even a fractional occupancy of this orbital combined with a large 
$\pi 1g_{9/2}$ population can bring a substantial raise of the collectivity.
The $\nu 2f_{7/2}$ (and to a lesser extent $\nu 3p_{3/2}$) ESPE are in the proximity of the rest of the shells as well, which
fuels the quasi-SU3 mechanism suggested in \cite{Sieja-zr}.  
The  neutron $1g_{9/2}$ is not shown as it is located 12 MeV below the $2d_{5/2}$. 
The necessity of higher intruder orbits to get the right degree of quadrupole correlations in the shell-model
framework was thus confirmed in Zr isotopes. 
On the contrary, the importance of the $\nu 1g_{9/2}$ in the deformation-driving process
could be rouled out based on those MCSM results.

\begin{figure}[H]
\begin{center}
\includegraphics[width=0.4\textwidth]{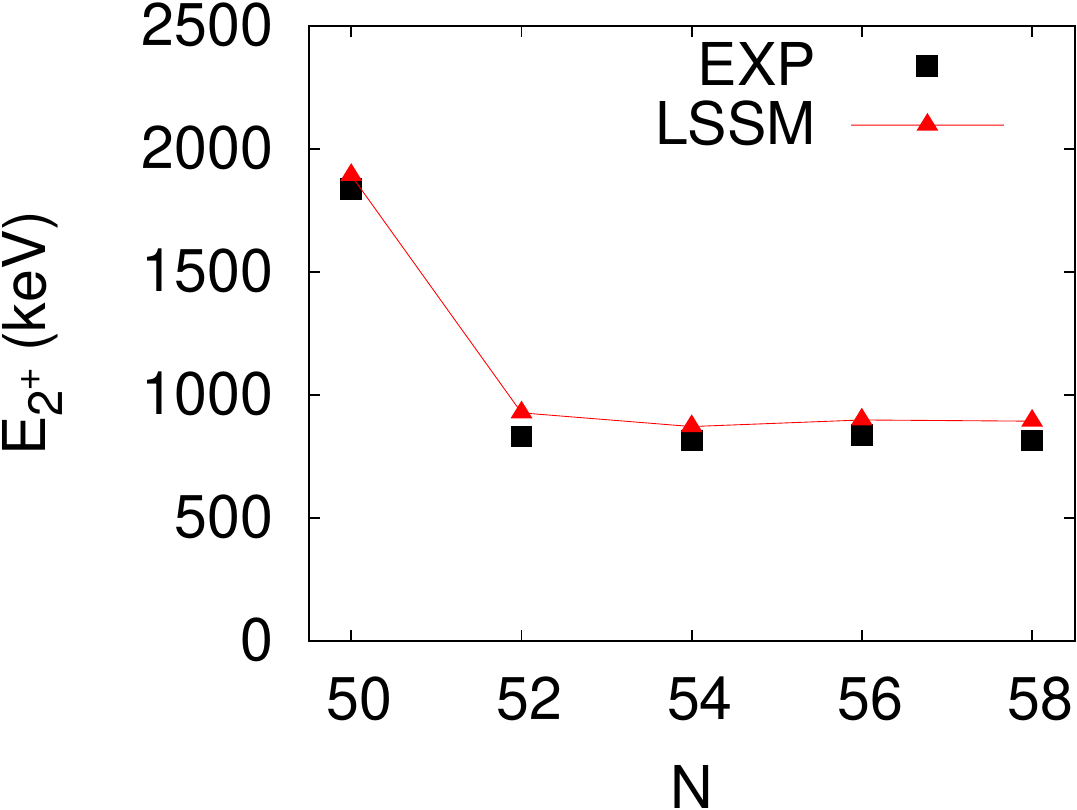}
\caption{$2^+$ excitation energy along the Sr chain: experiment vs present calculations.\label{fig-e2}}
\end{center}
\end{figure}

In the previous section the behavior of the $0^+$ states was discussed. 
The present model predicts  
a correct number of low energy $0^+$ states and is sufficient to describe the systematics of the majority 
of yrast and non-yrast low-spin states from $N=50$ to $N=58$ included,
as was shown in Ref. \cite{Urban-Sr2021}. 
For completeness, the $2^+$ excitation energies are shown in Figure \ref{fig-e2}.
The agreement is very good in the whole chain, with the maximum difference of only 90~keV. 
The calculation reproduces correctly the tendency without a shell 
closure effect at $N=56$. One should bear in mind that the $2^+$ energy raises in the neighbouring 
$^{96}$Zr which is well given by the same calculations \cite{Sieja-zr}.    

\begin{figure}[H]
\begin{center}
\includegraphics[width=0.4\textwidth]{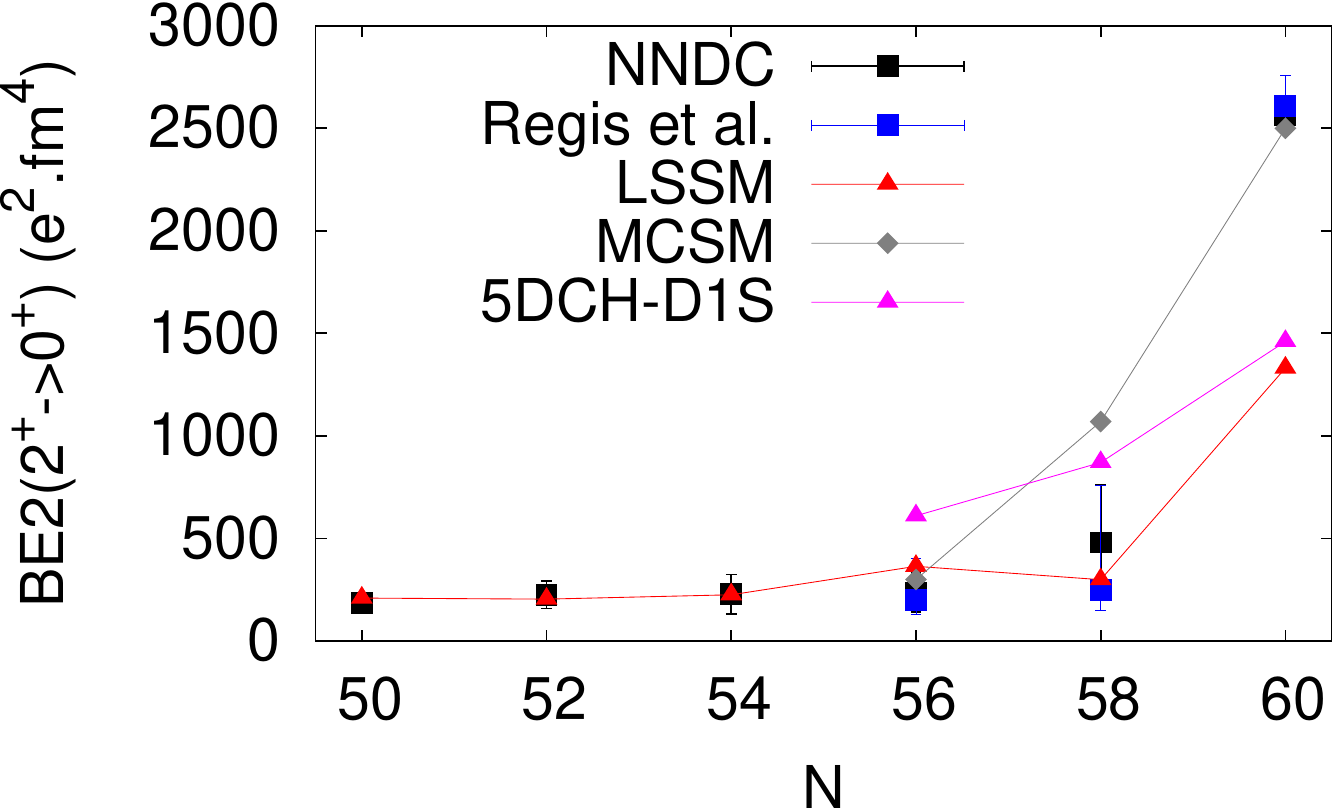}
\caption {$E2$ reduced transitions probabilities as obtained within the present framework
in comparison to available data and other models predictions. The LSSM value at $N=60$
is obtained in a fixed-configuration calculation, see text for details.
The MCSM and 5DCH-D1S values are those reported along with the data
in Ref. \cite{Regis-Sr}.\label{fig-be2}}
\end{center}
\end{figure}

In Figure \ref{fig-be2} the $B(E2; 2^+\rightarrow 0^+)$ transitions are shown
in comparison to experimental data and other models taken from \cite{Regis-Sr}.
In the calculations of 
the reduced transition probabilities $0.7e$ polarization charge was applied for both neutrons 
and protons. Note that in Ref. \cite{Sieja-zr} $0.8e$ charge was employed which appears to overestimate transitions 
in Sr isotopes. One could fine-tune further the proton and neutron charges for a better agreement with experiment 
on particular transitions. Nevertheless, the experimental errors are large in the transition region
and the major interest is to understand 
the relative differences in the magnitude of transitions between
the isotopes. 
Present calculations fit very well the data in lighter Sr isotopes. At $N=56$, the predicted value is close to that of MCSM
and indicates an increase of collectivity with respect to $N=54$. This comes from the large quadrupole matrix elements between
$2d_{5/2}$ and $3s_{1/2}$ orbitals which are both well occupied due to the lack of the shell closure at $N=56$.  
Such an increase is in contrast with the flat behavior 
of experimental values, though the shell-model ones still fall within the error bars. 
The $B(E2)$ from the 5DCH model with Gogny forces is three times larger
than experiment at $N=56$. 
In $^{96}$Sr, at $N=58$, the deviations grow: the available experimental values differ a lot one from another
though are consistent within the error bars.
The 5DCH model predicts more collectivity at $N=58$ than at $N=56$. 
MCSM gives the largest of all $B(E2)$ values presented: as mentionned earlier,
those calculations seem to predict the shape transition at a too low neutron number in Sr. Interestingly,
the present calculation falls a bit down towards the lower of the two experimental values.  
Wether this behavior is correct can be further debated - as deduced before from the spectroscopic factors calculations,
the ground-state wave-function of $^{96}$Sr may not be very accurate.  
More experimental and theoretical effort can still be done to provide a comprehensive picture of the
coexisting formes just before the shape change at $N=60$.

After the shape change takes place at $N=60$, the MCSM value matches very well the experiment while the 5DCH-D1S does not increase enough.
In spite of that, the latter calculation reproduces correctly a set of other transitions in the same nucleus
(see Refs. \cite{Regis-Sr, Clement2016}. In the case of present LSSM, 
only a fixed-occupancy calculation was performed in $^{98}$Sr:
The occupations of the spin-orbit partners $1g_{9/2}-1g_{7/2}$ 
were fixed to 4 and 6 particles, respectively, and seniority $\nu=12$ was allowed.  
The $2^+$ state in such a calculation is located at 284keV and the $B(E2)$ transition value, shown
in Fig. \ref{fig-be2}, is 1331$e^2fm^4$. As seen, having an increased occupancy of these orbits (of a similar order as resulting
fully mixed MCSM calculations in Zr isotopes), leads to a great rise of the $B(E2)$ value compared to $N=58$ but still 
not sufficient to match the experiment. 
In the $0^+$ state computed without imposing occupancies (at $\nu=10$)
there is 0.45 particle in $1g_{9/2}$ and 0.9 particle in $1g_{7/2}$, only. The $B(E2;2^+\rightarrow0^+)$ 
transition is then one order of magnitude lower than in a fixed-occupation calculation. 
Clearly, the interaction consistent with the properties of lighter Sr isotopes and a large number of other nuclei in the 
region does not favor configurations with many particles in the SOP to take over. 
This was also the case of Zr isotopes studied in the same framework which confirms the deformation 
origin is the same in both chains. The immediate conclusion is that its description requires 
an extra mechanism to populate the $1g_{9/2}$ and $1g_{7/2}$ orbitals.
The second observation is that additional quadrupole collectivity is necessary, even if SOP are well occupied.
Both points can not to be satisfied without extending the valence space, 
testifying the crucial role of intruder orbitals in shaping nuclei in the region. 
It is now advisable to investigate further theoretically  
the shape coexistence around $N=60$ towards lighter-$Z$ nuclids. 
As in Kr and Se the drops of $2^+$ energies are not 
as pronounced as in Sr and Zr, it is of interest to uncover the origin of this difference 
and to track the evolution of the intruding orbits
with decreasing proton number.
  
\section{Conclusion \label{conc}}
The properties of Sr isotopes described in a valence space outside the $^{78}$Ni core were discussed.
Present calculations reproduce properly the excitation energies and wave functions of low-energy states
between the $N=50$ and $N=56$ and to some extent at $N=58$, as was previously the case
of Zr isotopes. Accounting fully for the abrupt shape change at $N=60$ appears impossible
without incorporation of orbitals from adjacent harmonic oscillator shells.

Compared to the Zr isotopes, the strontiums appear more challenging for the present shell-model description.
While the structure of odd nuclei seems well reproduced and understood, the even-even ones
reveal systematic differences with experiment. In particular, the energies of excited $0^+$ states
are underestimated, probably due to inaccuracies in pairing interactions of several orbitals involved.
The difficulty of reproducing the coexisting
structures in Sr isotopes to a great detail seem common to the available shell-model and other approaches.
On the other hand, it is clear from the current and previous calculations that the origin of deformation in Zr and Sr
isotopes should be the same. The famous atrraction mechanism between neutrons in $1g_{9/2}$
and protons in $1g_{7/2}$ appears insufficient to yield enough of collective enhancement in the $B(E2)$ values
at $N=60$. Additionally, the effective interaction reproducing properties of large number of nuclei in the region
does not favor configurations with highly occupied SOP to dominate the low-energy states. 
Thus the presence of intruder orbitals can not be neglected to describe the quadrupole collectivity in Sr and Zr with $N\ge58$. 
As the lighter-$Z$ isotopes do not reveal such abrupt changes in their structure when passing
$N=60$, it would be now of interest to provide a microscopic description of the shape coexistence
of those nuclei in the same theoretical framework.

In spite of the recent experimental progress, there are still missing ingredients that could help
to understand the structure of low-energy excitations in Sr and in neighbouring isotopes. 
As an example, it is expected that the neutron $1g_{7/2}$ orbital plays
an important role in driving the deformation in this region of nuclei but there are no
exprimental constraints permitting to verify its position close to $N=50$ and to follow its evolution
with the neutron number.
Searching experimentally for spherical states including 1p-1h excitations
to this orbit could be of interest for future theoretical developments.
Also, the understanding of the coexisting shapes before $N=60$
and the presence of the quantum shape transition in Sr isotopes could
be deepen through 2 neutron and $\alpha$ transfer reactions. 
This would clearly help to examine the pairing collectivity
of the low-energy $0^+$ excitations and elucidate the wave-function
decomposition in Sr isotopes.  

\externalbibliography{yes}                                                                                                                                               
\bibliography{kama.bib}


\end{paracol}
\end{document}